\documentstyle[12pt,epsf]{article}

\renewcommand{\bar}{\overline}
\newcommand{\newc}{\newcommand}
\newc{\shat}{\hat{s}}
\newc{\invpb}{\,\mbox{pb}^{-1}}
\newc{\gev}{\,\mbox{GeV}}
\newc{\mev}{\,\mbox{MeV}}
\newc{\gsim}{\lower.7ex\hbox{$\;\stackrel{\textstyle>}{\sim}\;$}}
\newc{\lsim}{\lower.7ex\hbox{$\;\stackrel{\textstyle<}{\sim}\;$}}
\newc{\beq}{\begin{equation}}
\newc{\eeq}{\end{equation}}
\newc{\bea}{\begin{eqnarray}}
\newc{\eea}{\end{eqnarray}}
\newc{\eg}{{\it e.g.}}
\newc{\ie}{{\it i.e.}}
\newc{\etal}{{\it et al}}
\newc{\eps}{\epsilon}
\def\PLB#1#2#3{Phys. Lett. {\bf B#1} (19#2) #3}

\def\PRD#1#2#3{Phys. Rev. {\bf D#1} (19#2) #3}
\def\PRL#1#2#3{Phys. Rev. Lett. {\bf#1} (19#2) #3}

\def\ZPC#1#2#3{Zeit. f\"ur Physik {\bf C#1} (19#2) #3}

\catcode`@=11
\long\def\@caption#1[#2]#3{\par\addcontentsline{\csname
  ext@#1\endcsname}{#1}{\protect\numberline{\csname
  the#1\endcsname}{\ignorespaces #2}}\begingroup
    \small
    \@parboxrestore
    \@makecaption{\csname fnum@#1\endcsname}{\ignorespaces #3}\par
  \endgroup}
\catcode`@=12
\begin{document}

\begin{titlepage}
\begin{flushright}
{\rm
IASSNS-HEP-97-04\\
hep-ph/9703299\\
March 1997\\
}
\end{flushright}
\vskip 2cm
\begin{center}
{\Large\bf Comments on the high-$Q^2$ HERA anomaly\footnote{Research
supported in part by DOE grant DE-FG02-90ER40542, and by the W.~M.~Keck
Foundation. Email: {\tt babu@sns.ias.edu, kolda@sns.ias.edu,
jmr@sns.ias.edu, wilczek@sns.ias.edu}}}
\vskip 1cm
{\large
K.S.~Babu,
Christopher Kolda,
John March-Russell\\
and\\
Frank Wilczek\\}
\vskip 4pt
{\large\sl School of Natural Sciences,\\ Institute for Advanced Study,\\
Princeton, NJ, USA~~08540\\}
\end{center}
\vskip .5cm
\begin{abstract}
Taking the reported high-$Q^2$ anomaly at HERA as a signal of new physics,
we show that several independent considerations point 
towards $s$-channel leptoquark production
as the most attractive interpretation.
We argue that even this option is highly 
constrained by flavor-changing processes:
the couplings must be accurately diagonal in the quark and lepton mass
eigenstate basis and should preserve individual quark and lepton family 
numbers. We propose a dynamical mechanism that might produce
this pattern; it has distinctive experimental consequences.
\end{abstract}
\end{titlepage}
\setcounter{footnote}{0}
\setcounter{page}{1}
\setcounter{section}{0}
\setcounter{subsection}{0}
\setcounter{subsubsection}{0}

%%%%%%%%%%%%%%%%%%%%%%%%%%%%%%%%%%%%%%%%%%%%%%%%%%%%%%%%%%%%%%%%%%%%%%%

Recently, the H1~\cite{H1} and ZEUS~\cite{zeus} collaborations,
which have been studying $e^\pm p$
collisions on the HERA ring since 1993, announced an anomaly in
high-$Q^2$ $e^+p$ collisions. Using a combined accumulated luminosity of
$34.3\invpb$ in $e^+p\to eX$ mode at $\sqrt{s}=300\gev$, the two
experiments have
observed 24 events with $Q^2>15000\gev^2$ against a Standard Model (SM)
expectation of $13.4\pm1.0$, and 6 events with $Q^2>25000\gev^2$ against
an expectation of only $1.52\pm0.18$.
If we take these reports at face value, there seems to be little
possibility of explaining the anomaly within the Standard Model.
In particular, the
high-$Q^2$ events are clustered at Bjorken-$x$ values near
0.4 to 0.5; the quark distribution functions at such large $x$ are
well-measured
at lower $Q^2$, and QCD allows for a very precise scaling to the observed
$Q^2$. (By contrast, the superficially similar reported anomaly in the
high-$E_T$ dijet cross-section at FNAL~\cite{dijet}
can plausibly be ascribed to uncertainties in
the gluon distribution functions.)
The peaked distribution
of these high-$Q^2$ events in $x$ is also notable.
If the new physics is assumed to be
an $s$-channel resonance, this distribution
translates into a mass determination. For both H1 and ZEUS two independent
determinations are possible, depending on whether $x$ is calculated from
the double-angle or electron methods. For the seven events selected for
special study by H1 we find these two methods give $M_{2\alpha}=202\pm14\gev$
and $M_e=199\pm8\gev$ respectively. Similarly, for the five events
selected by ZEUS, $M_{2\alpha}=231\pm16\gev$ and $M_{e}=219\pm12\gev$ .

HERA is capable of running in two modes: $e^-p$ and $e^+p$. In the former mode
H1 and ZEUS have accumulated $1.53\invpb$ of data but have observed
no statistically significant deviations from the SM. Further, the experiments
differentiate between final state $eX$ and $\nu X$, where the neutrino is
detected through its missing $p_T$.
The anomaly discussed above is in the neutral
current (NC) reaction $e^+p\to eX$ channel. H1 has also announced its
findings in the $e^+p\to\nu X$ charged current (CC) channel. They find
3 events at $Q^2>20000\gev^2$ with an expectation of $0.74\pm0.39$, but
no events with $Q^2>25000\gev^2$. ZEUS has not announced its CC data.
Compared to the NC channel, the CC signal is considerably less statistically
significant, and we therefore regard the question of whether new physics
in this channel is required as open.

Both detectors at HERA are highly
efficient at tagging muons. Apart from one such event in the 1994 H1
data~\cite{muevent} (which occurs in a very different kinematic region
from the events under consideration here), neither experiment has
evidence for final state muons.

Reprising: H1 and ZEUS see an excess of high-$Q^2$ NC events in $34.3\invpb$
of $e^+p$ data, with no corresponding anomaly in $1.53\invpb$ of $e^-p$ data.
There is no clear evidence for or against a similar anomaly in the CC
mode. There is no signal for final state $\mu$'s. There is evidence that the
events seen are peaked at $x\sim0.45$--$0.55$, but there is some disagreement
between the two experiments concerning the central $x$ value at the peak. If
the signal is real, and not a statistical fluctuation, there appears to be
no way to change the parameters of the SM to explain it. Let us entertain the
hypothesis that this data represents a genuine signal,
and investigate the consequences.

~

{\it New Physics.} What kind of new
physics could be responsible for the HERA anomaly? We consider only physics
that can be described by an
$SU(3)\times SU(2)\times U(1)$-invariant effective Lagrangian using
perturbation theory.\footnote{Thus our remarks do not apply to proposals
such as that of Ref.~\cite{adler}.}

There are 3 kinematic channels through which the $ep$ scattering might be
occurring: in the $s$-channel through a resonance with non-zero lepton
and baryon numbers (a ``leptoquark''), in the
$u$-channel again via a leptoquark,
or in the $t$-channel via a particle
without lepton or baryon number (this includes Higgs or $Z'$ exchange or
contact interactions).
We will discuss the $t$ and $u$-channels first.

If the intermediate state is a $t$-channel scalar or vector, then there are
two kinematic limits in which very different constraints play a role. For
a light $Z'$ (the results hold equally
well for a scalar with a suitably rescaled coupling), the CDF and D0
experiments provide strong
bounds from searches for $\bar qq\to Z'\to e^+e^-, \mu^+\mu^-$.
Assuming significant
branching ratios of the $Z'$ to leptons, they rule out $Z'$ masses
below about $650\gev$~\cite{zprime}.
For masses greater than this bound, the $Z'$ can
be integrated out and one can work in terms of effective 4-fermion contact
interactions. By convention, one writes the operators:
\beq
{\cal L}=\sum_{i,j=L,R}
\frac{4\pi}{(\Lambda^q_{ij})^2}\,\eta^q_{ij}\,
(\bar e_i\gamma_\mu e_i)(\bar q_j\gamma^\mu q_j)
\eeq
where $q=u,d$ and 
$\eta^q_{ij}=\pm1$. Thus there are 8 different contact interactions,
times 2 signs of $\eta$.

To analyze how well each operator fits the observed data at HERA,
we perform a standard $\chi^2$ analysis with one free parameter, namely
$\Lambda^q_{ij}$. For the fit, we use the combined
bin-by-bin data of ZEUS and H1 for bins starting at $Q^2$=(10000, 15000, 20000,
25000, 30000, 35000)$\gev^2$ respectively, with each bin (except the last)
being $5000\gev^2$ wide; the last bin goes from $Q^2=35000\gev^2$ to
$Q^2=s=90000\gev^2$. Bin by bin, ZEUS and H1 combined see (29, 14, 4, 2,
2, and 2) events. 

For each operator and each choice of $\Lambda^q_{ij}$ we calculated the 
number of events expected in each $Q^2$ bin at HERA. H1 and ZEUS claim
efficiences approaching 80\% in the $Q^2$ region of interest, and our own
calculations of the SM and new physics cross-sections were 
scaled appropriately. For the parton-level calculation, we generalized the
results of Ref.~\cite{contact}\ to contact interactions. The parton-level
calculations were then folded into the CTEQ3M parton distribution 
functions~\cite{CTEQ}\ to get the physical event rates.

The Standard Model, with 6 degrees of freedom has a
$\chi^2/dof$=4.0, which is disfavored at greater than $99.9\%$ C.L.
For the contact terms, we demand a fit better than $2\sigma$ (95\% C.L.)
to the data, which for $6-1=5$ degrees of freedom, translates to
$\chi^2/dof<2.2$. For each operator, we find upper bounds on $\Lambda^q_{ij}$
as given in Table~\ref{table:contact}. OPAL has recently announced
95\% C.L. lower bounds on the scales $\Lambda$ of each of these
operators~\cite{OPAL}. Given these
results, only the $\Lambda^u_{LR}$ and $\Lambda^u_{RL}$ operators (with
$\eta=+1$) can have sufficient strength to explain the HERA data, and
simultaneously satisfy the OPAL bounds.
Of course, neither allowed operator will produce a peaked distribution in $x$.
%%%%
\begin{table}
\centering
\begin{tabular}{c|cc|cc}
~ & \multicolumn{2}{c}{$q=u$} & \multicolumn{2}{c}{$q=d$} \\
$ij$ & $\eta=+1$ & $\eta=-1$ & $\eta=+1$ & $\eta=-1$ \\
\hline
LL & -- & 0.8 & -- & -- \\
RR & -- & 0.8 & -- & -- \\
LR & 2.9 & 1.2 & 1.2 & 1.5 \\
RL & 2.4 & 1.4 & 1.4 & 1.3
\end{tabular}
\caption{95\% C.L. upper bounds on the $\Lambda^q_{ij}$ in TeV. Dashed entries
are operators which do not fit the HERA data for any $\Lambda$.}
\label{table:contact}
\end{table}
%%%%

In any case, there is another strong constraint on these interactions, arising
from atomic parity violation (APV) measurements.
Experiments using cesium ($Z=55$, $N\simeq78$)
have found its weak charge to be $Q_W^{\rm exp}({\rm Cs})=-71.04\pm1.81$, while
the SM predicts $Q_W^{\rm SM}({\rm Cs})=-73.12\pm0.09$~\cite{pdb}.
A contact interaction shifts $Q_W$ by an amount~\cite{limits}:
\beq
\Delta Q_W=-2\left[C_{1u}(2Z+N)+C_{1d}(2N+Z)\right]
\eeq
where
\beq
C_{1q}=\frac{\sqrt{2}\pi}{G_F}\left(\frac{\eta^q_{RL}}{(\Lambda^q_{RL})^2}-
\frac{\eta^q_{LL}}{(\Lambda^q_{LL})^2}-\frac{\eta^q_{LR}}{(\Lambda^q_{LR})^2}
+\frac{\eta^q_{RR}}{(\Lambda^q_{RR})^2}\right).
\eeq
For any one operator with $\Lambda\lsim$ 3 TeV, $\Delta Q_W\approx\pm20$,
which is decisively excluded by the APV experiments.
Logically, we cannot exclude the possibility of a conspiracy,
with cancellations among individually large contributions from
several operators (this can be natural however for contact interactions
generated by a parity-conserving
gauge interaction), but clearly these contact
terms provide at best a marginally consistent parameterization
for understanding the HERA data.

A $u$-channel leptoquark $\Phi$, in the heavy mass limit, goes over into the
same types of contact interactions just discussed.
But in this case an additional problem for light leptoquarks presents itself 
within the HERA data. Given $e^+p$ scattering in the $u$-channel
with a valence (or sea) quark, by crossing and charge-conjugating the Feynman
diagram, one arrives at $e^-p$ scattering in the $s$-channel off the same
valence (or sea) quark. Therefore $\sigma(e^-p)\simeq(M_\Phi/\Gamma_\Phi)
\sigma(e^+p)$ due to resonant production of $\Phi$ in $e^-p$ scattering,
where $\Gamma_\Phi$ is the leptoquark width:
\beq
\Gamma_{{\rm scalar~LQ}}
=n\frac{\lambda^2}{16\pi}M_{LQ},\quad\quad\quad\quad
\Gamma_{{\rm vector~LQ}}=n\frac{\lambda^2}{24\pi}M_{LQ},
\eeq
where $n$ is the number of available leptoquark decay channels.
Given typical values $\Gamma_\Phi\sim10\mev$, then
even with its much smaller integrated luminosity, this represents a signal of
about $10^3$ events in $e^-p$ in the current data. 
Thus $u$-channel leptoquark exchange
cannot be responsible for the HERA anomaly.

Thus $s$-channel production is the most attractive option. It is, moreover,
consistent with the peaked distribution in $x$ observed
at HERA. In this channel an important question arises, whether
the $e^+p$ scattering is off valence or sea quarks. If the scattering
were off sea quarks, then by charge-conjugating the relevant diagram, one
would arrive at $e^-p$ scattering off valence quarks, also resonant.
At $x\simeq0.5$, the valence and sea parton
distribution functions come in the approximate ratio $u_v\approx 4d_v\approx
200\bar u_s=200\bar d_s$. (Recall that as $x\to1$, $u_v(x)\approx
2d_v(x)/(1-x)$.) So for every one anomalous event
observed in the $e^+p$ mode,
H1 and ZEUS should have found between 2 to 9 such events in the $e^-p$ mode
(depending on whether the coupling is to $u$ or $d$-quarks), despite
the small integrated luminosity. Since no such excess has been
observed, the $e^+p$ anomaly must be described by
$s$-channel production of a leptoquark coupling to valence quarks in
the proton.

~

{\it Scalar Leptoquarks.}
For scalar leptoquarks, there are only a few operators which are
gauge-invariant, renormalizable and couple to both electrons and quarks.
They are listed in Table~\ref{table}. Also shown are the quantum
numbers of the leptoquark, $\Phi$, under $SU(3)\times SU(2)\times U(1)_Y$.
In the column ``CC'' is indicated 
whether or not the operator leads to CC events
at HERA in addition to the NC. In the last column (``mode'') is listed the
dominant production mode for the given leptoquark. We already argued that
the events at HERA must be coming from $s$-channel scattering off of valence
quarks in the proton; such operators are listed as ``$e^+$'' mode in the table.
The remaining ``$e^-$'' operators would have been already detected in HERA's
$e^-p$ data given the size of coupling necessary to explain the anomaly
in the $e^+p$ data.

%%%%%%%%%%%%%%%%%%%%%%%%%
\begin{table}
\centering
\begin{tabular}{c|ccc|cc}
Operator & SU(3) & SU(2) & U(1) & CC & mode \\ \hline
$L_iQ_j\Phi\eps^{ij}$ & $\bar3$ & 1 & ${1/3}$ & Yes & $e^-$ \\
$L_iQ_j\Phi^{ij}$ & $\bar3$ & 3 & ${1/3}$ & Yes & $e^-$ \\
$L_i\bar u\Phi_j\eps^{ij}$ & 3 & 2 & ${7/6}$ & No & $e^+$ \\
$L_i\bar d\Phi_j\eps^{ij}$ & 3 & 2 & ${1/6}$ & No & $e^+$ \\
$\bar e Q_i\Phi_j\eps^{ij}$ & $\bar3$ & 2 & $-{7/6}$ & No & $e^+$ \\
$\bar e\bar u\Phi$ & 3 & 1 & $-{1/3}$ & No & $e^-$ \\
$\bar e\bar d\Phi$ & 3 & 1 & $-{4/3}$ & No & $e^-$
%\\ \hline
%$\bar L_i\gamma_\mu Q_j\Phi^\mu\eps^{ij}$ & $\bar3$ & 1 & $-\frac{2}{3}$
%& Yes & $e^+$ \\
%$\bar L_i\gamma_\mu Q_j\Phi^{\mu ij}$ & $\bar3$ & 3 & $-\frac{2}{3}$
%& Yes & $e^+$ \\
%$\bar L_i\gamma_\mu \bar u\Phi^\mu_j\eps^{ij}$ & 3 & 2 & $-\frac{1}{6}$
%& No & $e^-$ \\
%$\bar L_i\gamma_\mu\bar d\Phi^\mu_j\eps^{ij}$ & 3 & 2 & $\frac{7}{6}$
%& No & $e^-$ \\
%$e\gamma_\mu Q_i\Phi^\mu_j\eps^{ij}$ & $\bar3$ & 2 & $\frac{5}{6}$
%& No & $e^-$ \\
%$e\gamma_\mu\bar u\Phi^\mu$ & 3 & 1 & $\frac{5}{3}$ & No & $e^+$ \\
%$e\gamma_\mu\bar d\Phi^\mu$ & 3 & 1 & $\frac{2}{3}$ & No & $e^+$
\end{tabular}
\caption{List of scalar leptoquark operators. For each operator, the
$SU(3)\times SU(2)\times U(1)_Y$ quantum numbers of the leptoquark,
$\Phi$, are shown. The fifth column
indicates whether HERA should find CC events, and the final column lists the
mode in which HERA should dominantly produce the given leptoquark.
$Q$ and $L$ represent $SU(2)$ doublet quarks and leptons, while $\bar e$,
$\bar u$ and $\bar d$ are $SU(2)$ singlets.}
\label{table}
\end{table}
%%%%%%%%%%%%%%%%%%%%%%%%%

Note that if a leptoquark couples significantly to both LH and RH quarks,
it generates effective 4-Fermi operators of the
form $\bar{u_R}d_L\bar{e_R}\nu_L$.  These lead, for example, to
helicity-unsuppressed $\pi\to e\nu$ decays, which are severely
constrained by experiment~\cite{limits}. Among the operators relevant
at HERA, this means that we may not identify $\Phi_{eQ}=\Phi^*_{Lu}$
(where the subscripts indicate the SM fermions to which the
leptoquark couples )
despite their identical
quantum numbers. Alternatively, for a $\Phi(3,2,7/6)$ we must
forbid one or the other of its $L\bar u\Phi$ or $\bar eQ\Phi^*$ couplings.

Of the 7 leptoquark operators listed in Table~\ref{table}, only 3 are
consistent with the HERA data as we interpret it: $\Phi_{Lu}$, $\Phi_{Ld}$
and $\Phi_{eQ}$. 
All three are $SU(2)$ doublets, and we will assume throughout the rest of
this analysis that the member coupling to electrons is the lighter so that
it does not decay to the component coupling to neutrinos and a real or 
virtual $W$. Such a decay would have a very different signature at
HERA. We also
note that the leptoquark which can appear in a ${\bf\bar5}$ of $SU(5)$ is not
among the three possible states useful for explaining the HERA data; however,
the $\Phi_{Ld}$ leptoquark can fit into a {\bf 10} of $SU(5)$.

Notice that all scalar leptoquarks which would have
given a CC signal have been ruled out already by lack of NC events in $e^-p$
mode. Thus for scalar leptoquarks, we conclude that there will
be no CC signal at HERA. Conversely, if a CC signal is seen at HERA, a
scalar leptoquark interpretation is strongly disfavored.

We performed a similar $\chi^2$ analysis here to the one described above for
contact interactions, except there are now 2 independent variables:
$M_{LQ}$ and $g_{LQ}$. Relevant cross-section calculations have been
performed in Ref.~\cite{buchmuller}.
Here again the APV measurement can strongly constrain
the range of parameters allowed. For scalar leptoquarks one finds
\beq
\Delta Q_W^{\rm LQ }=-2\left(\frac{g_{LQ}/M_{LQ}}{g_W/M_W}\right)^2
(\delta_Z Z+\delta_N N).
\eeq
For the leptoquarks of interest, the values of $(\delta_Z,\delta_N)$
are: (-2,-1) for $\Phi_{eQ}$, (2,1) for $\Phi_{Lu}$, and (1,2) for
$\Phi_{Ld}$. For a given mass and coupling, the constraint on $\Phi_{eQ}$
will be weaker than for the others, since it shifts $Q_W$ in the
direction of experiment.

\begin{figure}[t]
\centering
\epsfxsize=4in
\hspace*{0in}
\epsffile{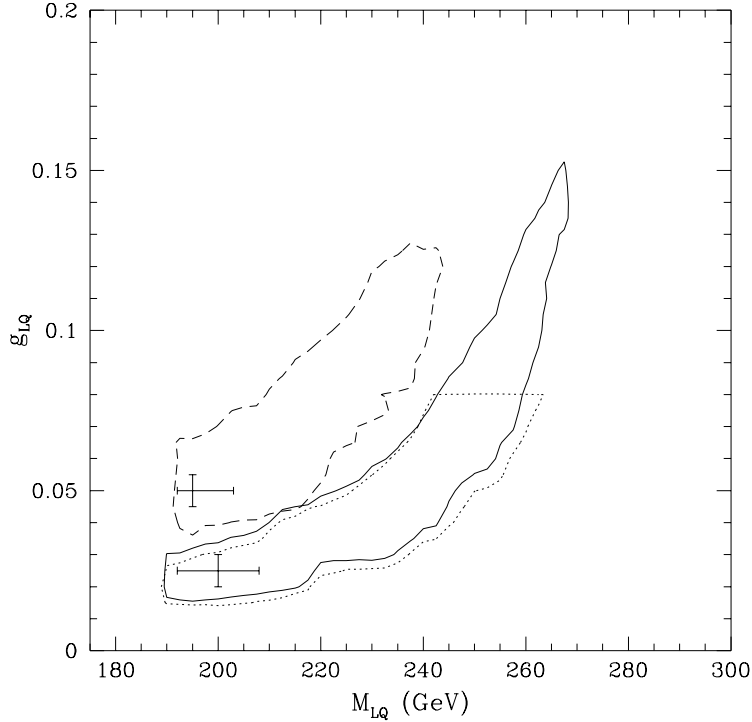}
\caption{95\% C.L. regions in the $(M_{LQ}, g_{LQ})$ plane derived
from fitting to the HERA $Q^2$ distribution and APV in cesium. The three
scalar leptoquark operators are shown: $\Phi_{Lu}$ (solid), $\Phi_{eQ}$
(dotted), and $\Phi_{Ld}$ (dashed). For $\Phi_{Lu}$ and $\Phi_{Ld}$ the
best fit values for $M_{LQ}$ and $g_{LQ}$ are shown with $1\sigma$
error bars. For $\Phi_{eQ}$, couplings above 0.08 are excluded by a
combination of rare decay constraints.} 
\label{fig:chi2}
\end{figure}

Given the $6+1-2=5$ degrees of freedom, we explore the range of $(M_{LQ},
g_{LQ})$ which give $\chi^2/dof<2.2$ (\ie, 95\% C.L.). The 95\% confidence
regions are shown in Figure~\ref{fig:chi2}\ for each of $\Phi_{eQ}$,
$\Phi_{Lu}$, and $\Phi_{Ld}$.
%%%%%%%%%%%%%%%%%%%%%%%%%%%%%%%%%%%%%%%%%%%%%%%%%%%%%%%%%%%%%%%%%%%
%%%%%%%%%%%%%%%%%%%%%%%%%%%%%%%%%%%%%%%%%%%%%%%%%%%%%%%%%%%%%%%%%%%
For $\Phi_{Lu}$ and $\Phi_{Ld}$, the leptoquark couplings are constrained to 
be small by the APV bounds discussed above.
For $\Phi_{eQ}$, we assume that the members of the leptoquark doublet are
nearly degenerate and both are produced at HERA; if one or the other is
significantly heavier, then the best fit values as far as HERA is concerned
go over to those of $\Phi_{Lu}$ or $\Phi_{Ld}$, as appropriate.
It may seem that in the combined HERA + APV analysis for $\Phi_{eQ}$,
large values of the corresponding coupling,
$g_{LQ}$, still provide good fits since they improve agreement of theory
with the APV data. However, as we discuss later, a combination of rare
decay constraints coming from $K_L \to \pi^0\ell^+\ell^-$ and
$D^0\to \pi^0\ell^+\ell^-$ exclude values of $g_{LQ}>0.08$ for
this operator for masses in the interesting range. 

We also calculate ``$1\sigma$'' bounds on $M_{LQ}$ and $g_{LQ}$ by
demanding $\chi^2\leq\chi^2_{\rm min}+1$. These bounds are not true
$1\sigma$ bounds since the iso-$\chi^2$ lines are far from elliptical 
but we consider the bound indicative of where the best fit values fall.
For the $\Phi_{Lu}$ and $\Phi_{Ld}$ leptoquarks, one finds:
\beq
\begin{array}{lll}
\Phi_{Lu}: & M_{LQ}=(200\pm8)\gev, & g_{LQ}=0.025\pm0.005 \\
\Phi_{Ld}: & M_{LQ}=(195^{+8}_{-3})\gev, & g_{LQ}=0.05\pm0.005
\end{array}
\eeq
A similar bound on the $\Phi_{eQ}$ leptoquark is difficult to determine. The
$\chi^2$ is small over a wide range of masses ($190\gev<M_{LQ}\lsim
260\gev$) and couplings ($0.01<g_{LQ}\lsim 0.08$), and once again it is only
the $x$-distribution at HERA which can further constrain them.

It is noteworthy that the fits have no information put
into them {\it a priori} about the observed distributions in $x$. Nevertheless,
for the $\Phi_{Lu}$ and $\Phi_{Ld}$ leptoquarks,
the best fit values to the shape of the $Q^2$ distribution are at $M_{LQ}$
consistent with the $M_{LQ}=\sqrt{xs}$ extracted from the HERA data.
We consider this
an additional piece of evidence in favor of an $s$-channel interpretation of
the HERA results.

~

{\it Vector Leptoquarks.}
The HERA data, considered by itself, can be similarly well described
by an $s$-channel exchange of a {\it vector} leptoquark with
$M_{LQ}\approx 200\gev$ and coupling $\sqrt{2}$ smaller than its corresponding
scalar leptoquark. However, such states are disfavored by an analysis
of Run I data taken by the D0 collaboration
at Fermilab, as we now argue.  First, note that D0 has reported a bound
on 1st generation 
scalar leptoquarks of 175~GeV at the 95\% confidence level~\cite{D0}.
This corresponds to a bound on the leptoquark production cross-section
(including the effect of cuts and efficiencies) of
$\sigma\lsim 0.4\,$pb~\cite{pakvasa}.  Now, the production cross-section for
a pair of vector leptoquarks,
$\Phi_\mu$, at the Tevatron has a model dependence
that stems from the possibility of a coupling
$\kappa\Phi_\mu^\dagger G^{\mu\nu} \Phi_\nu$ to the gluon field strength
tensor $G^{\mu\nu}$.  (There is an insignificant contribution to the
production cross section from the leptoquark Yukawa couplings given
the determination of their size from the HERA data.)
It can be seen from the analysis of Hewett,
\etal~\cite{hewett}\ that the contribution from $\kappa$ destructively
interferes most strongly with the process
$q\bar{q} \rightarrow g \rightarrow \Phi^\dagger\Phi$ at
values $\kappa\approx -0.5$.  For this most conservative value, the
production cross-section for pairs of vector leptoquarks at the Tevatron
exceeds 0.4~pb for masses $M_{LQ} < 215$~GeV~\cite{hewett}.  Thus such
masses are excluded at the 95\% confidence level.  If on the other
hand we take the value of $\kappa$ to lie in the range (0.0--1.0), then
masses $M_{LQ} \lsim 240\gev$ are excluded.  We therefore view vector
leptoquarks as only a marginally consistent possibility at best.

~

{\it Flavor Violation.}
The three effective interactions of scalar leptoquarks
that we have identified as capable of explaining
the HERA data are
\begin{eqnarray}
{\cal O} &=& \lambda_{ij}L_i \Phi d^c_j
= \lambda_{ij}(\nu_i d^c_j \phi^{-1/3} - e_i d^c_j \phi^{2/3}) \nonumber \\
{\cal O}' &=&\lambda'_{ij} L_i \Phi' u^c_j = \lambda'_{ij}(\nu_i u^c_j
\phi^{2/3} -e_i u^c_j \phi^{5/3}) \nonumber \\
{\cal O}'' &=& \lambda''_{ij}Q_i e^c_j \Phi'' = \lambda''_{ij}(u_i e^c_j
\phi^{-5/3} -V_{ik}d_k e^c_j \phi^{-2/3})~. \nonumber
\end{eqnarray}
(Lowercase $\phi^Q$'s correspond to the components of the leptoquark $SU(2)$
doublets of electric charge $Q$.)
Here the operators are written in the mass eigenbasis, where
$i,j$ are the generation indices and $V$ in ${\cal O}''$
is the CKM matrix.   The flavor structure of these Yukawa matrices
$(\lambda, \lambda',\lambda'')$ is severely constrained by
limits from rare processes (for earlier work in this regard, see
\cite{limits}).

To be definite take the operator ${\cal O}$ first.  If ${\cal O}$ explains
the HERA data, $\lambda_{11} \approx 0.05$ is needed.  The other
elements of $\lambda$ are then constrained as follows.   The decay
$K_L \rightarrow e^+e^-$ occurs at tree-level by the exchange of the
$\phi^{2/3}$ leptoquark.  However, the rate is strongly suppressed by a
helicity  factor $(m_e^2/m_K^2)$ and therefore does not give any
useful limit.  On the other hand, the decay $K_L \rightarrow
\pi^0 e^+e^-$, which arises from the same tree-level quark
diagram, is helicity unsuppressed.
The limit on this decay, $Br(K_L \rightarrow \pi^0 e^+ e^-) \le 4.3 \times
10^{-9}$, translates into a bound of $\lambda_{11}\lambda_{12} \le
1.5 \times 10^{-4} (M_{\phi^{2/3}}/200\gev)^2$.  For $\lambda_{11}
\approx 0.05$, this means $\lambda_{12} \le 3 \times 10^{-3}$.  Such
a hierarchy pattern seems quite unusual; it deviates considerably
from the pattern of Yukawa couplings that we have become used to in
the quark sector.

The doublet partner $\phi^{-1/3}$ of $\phi^{2/3}$ must be nearly degenerate
with it, else their mass--splitting will contribute to
the $\rho$--parameter excessively.  The tree--level exchange of
$\phi^{-1/3}$ causes the decay $K^+ \rightarrow \pi^+ \nu \overline{\nu}$.
The experimental limit $Br(K^+ \rightarrow \pi^+ \nu \overline{\nu}) \le
2.4 \times 10^{-9}$ leads to the bound $\lambda_{i1}\lambda_{2j} <
1.2 \times 10^{-4}(M_{\phi^{-1/3}}/200\gev)^2$.  One again sees a
severe limit on the second generation couplings.

The leptoquarks $\phi^{2/3}$ and $\phi^{-1/3}$
mediate the process $\mu \rightarrow e + \gamma$ through
one loop induced magnetic moment interactions.  The present limit on the
branching ratio, $Br(\mu \rightarrow e+ \gamma) < 4.9 \times 10^{-11}$, implies
that $\lambda_{11}\lambda_{21} < 1.6 \times 10^{-4} (M_\Phi/200\gev)^2$.
Similarly, the decay $K_L \rightarrow \mu e$ can occur at
tree level if both $\lambda_{11}$ and $\lambda_{22}$ are nonzero.  The limit
on this decay, $Br(K_L \rightarrow \mu e < 3.3 \times 10^{-11})$,
translates into a bound of $\lambda_{11}\lambda_{22}
< 2.8 \times 10^{-6} (M_\Phi/200\gev)^2$. Evidently it is not enough that
the couplings are flavor diagonal; individual quark and lepton number
must be preserved.

There are analogous constraints on the coupling $\lambda'_{ij}$ of
${\cal O}'$ coming from $D^0 \rightarrow \pi^0 e^+e^-$, etc, but
they are weaker numerically, by about a factor of 5, in the Yukawa
couplings.  The couplings $\lambda''_{ij}$ in ${\cal O}''$
have the interesting feature that one
cannot simultaneously suppress $D^0 \rightarrow \pi^0e^+e^-$ and
$K_L \rightarrow \pi^0 e^+e^-$.  
Suppressing the more strongly constrained mode $K_L \to \pi^0 e^+e^-$
leads to a limit on the individual {\sl diagonal} Yukawa
coupling $|\lambda_{11}''|^2 < 4 \times 10^{-3} (M_{\Phi''}/200\gev)^2$
from $D^0 \to \pi^0e^+e^-$. This is the bound imposed on the HERA
+ APV fit in Figure~\ref{fig:chi2}\ for the $\Phi_{eQ}$ operator.  
There is also a stringent limit on the product $(\lambda''^T V)_{11}
(\lambda''^T V)_{12}^*$ from $K_L\to \pi^0 e^+e^-$ decay.  In particular,
if only $\lambda''_{11}$ is nonzero (in this basis), the constraint
$|\lambda_{11}''|^2 < 6.8 \times 10^{-4} (M_{\Phi''}/200\gev)^2$, will
nearly exclude the possibility of explaining the HERA data with this operator.

Additional constraints arise from box diagrams which
induce $K^0-\overline{K}^0$ mixing, $D^0-\overline{D}^0$ mixing, etc,
but these are less severe than the ones quoted above, essentially
because they scale as $\lambda^4$ in the amplitude.  One process
worth noting is the box diagram contribution to $K_L \rightarrow
\pi^0\ell^+\ell^-$ and $D^0 \rightarrow \pi^0 \ell^+\ell^-$ with the exchange
of one $W^\pm$ and one leptoquark.  Such a diagram
directly constrains the diagonal leptoquark coupling, although
the limits are rather weak: $|\lambda_{22}|^2 \le 0.06 (M_\Phi/200\gev)^2$
from $K_L\to \mu^+\mu^-$.

~

{\it Conclusion: A Model.}
Limits arising from flavor physics provide very
severe constraints on any model containing leptoquark degrees of freedom
capable of explaining the HERA data. In the absence of a general principle,
it strains credulity to postulate the required large number of
tiny parameters in a fundamental theory.
An adequate interpretation of the HERA
data must address this question.  Neither the direct
introduction of fundamental leptoquark degrees of freedom, nor
other possibilities such as $R$-parity violating couplings of
MSSM squarks~\cite{rparity}, naturally explains this
remarkable flavor structure;
namely that the leptoquarks must, to a very high degree of accuracy,
couple diagonally to the {\sl mass}-eigenstate quarks and leptons, and
conserve individual quark and lepton numbers.
Clearly what is required is that the underlying physics responsible
for the HERA anomaly is universal in character with regard to the
generations.

The natural implementation of this
universality is a gauge principle, with the three generations
having identical charge assignments.  We have seen
that the HERA phenomenology disfavors $t$-channel $Z'$ exchange
as an explanation compared to the production of an $s$-channel
resonance.  We are therefore led to consider the possibility that a
new strong short-ranged gauge interaction leads to the formation of
relatively light leptoquark quasi-bound states.  Moreover, if the unification
of couplings observed in the MSSM is to have any chance of being
maintained, this gauge interaction must commute with the Standard
Model interactions.  As an example consider an additional $U(1)$ interaction
at a strong coupling fixed point, spontaneously broken at the TeV scale.
While one cannot reliably calculate the consequences of such an interaction,
a not unreasonable hypothesis for the dynamics of such a theory might be that
the low energy spectrum is saturated by resonances having small residual
interactions.  If the original gauge interaction commutes with $SU(5)$,
then the resonances automatically fill out several complete $SU(5)$
multiplets, each having different flavor quantum numbers.
More importantly, the generation independence of the
$U(1)$ gauge coupling implies that the masses and couplings
of the resonances (to free quarks and leptons) are to a good
approximation invariant under an $SU(3)$ symmetry that rotates the
generations.  This symmetry is only broken by effects that depend
on the usual Higgs Yukawa couplings, $h$, of the SM as $h^2$.
The flavor physics limits discussed above are easily
and naturally satisfied.

A model-independent prediction emerging from this scenario
is the existence of a large
multiplicity of very nearly degenerate resonances. This is to be 
expected from any model which preserves so carefully the $SU(3)$
flavor symmetries among the generations, as seems to be required
by the experimental flavor constraints. These states could be
discovered, or excluded, in future experiments at the Tevatron or LEP II.
There is already an interesting
constraint arising from FNAL data, since at minimum there are two
degenerate states containing an electron in combination with
either a 1st or 2nd generation quark.  From the D0 bound on
leptoquarks which couple to electrons~\cite{D0}\
the total production cross section to these states has
to be less than 0.4~pb, which translates to a limit of
$M_{LQ} > 190\gev$. Degenerate with these states will be leptoquarks
composed of $\mu q$ and $\tau q$, although the present bounds on such 
states are much weaker.

~

{\it Note added.}
While this paper was in preparation, several other
papers inspired by the high-$Q^2$ HERA anomaly appeared~\cite{new}.
Compared to other authors, we appear to place greater emphasis on the
unusual flavor structure required.

\section*{Acknowledgments}

We wish to thank M.~Barnett, I.~Hinchliffe, 
S.~Ritz, F.~Sciulli, S.~Treiman, and E.~Weinberg for useful
conversations.

\end{document}